# An Event-Driven Framework for Business Awareness Management

Babis MAGOUTAS[1], Dimitris APOSTOLOU[2] and Gregoris MENTZAS[1]

*[1]National Technical University of Athens,*
*Iroon Polytechniou 9 Zografou, Athens, 157 80 Greece,*
*[2]University of Piraeus, Karaoli & Dimitriou St. 80 Piraeus Greece 185 34*
*Email: elbabmag@mail.ntua.gr, dapost@mail.ntua.gr, gmentzas@mail.ntua.gr*

**Abstract:** Modern organizations need real-time awareness about the current business conditions and the various events that occur from multiple and heterogeneous environments and influence their business operations. Moreover, based on real-time awareness they need a mechanism that allows them to respond quickly to the changing business conditions, in order to either avoid problematic situations or exploit opportunities that may arise in their business environment. In this paper we present BEAM, an event-driven framework that enables awareness about the situations happening in business environments and increases organizations' responsiveness to them. We illustrate how BEAM increases the awareness of managers about the running business processes, as well as their flexibility by presenting a practical application of the framework in the transportation and logistics domain.

## 1. Introduction

To thrive in today's competitive market, organizations need to be agile, responding quickly to changing market conditions and exceeding customers' demands. Achieving business flexibility is a necessary condition for the business development of organizations, especially nowadays due to the global market downstream. However, business flexibility implies flexibility in the underlying ICT infrastructure and business processes.

A major challenge of current business process management solutions is to continuously monitor on-going activities in a business environment [1] and to respond to business events with minimal latency. Recent advancements in event-based systems and complex event processing [2] enable faster response to critical business events by efficiently processing many events occurring across all the layers of an organization and identifying the most meaningful ones within heterogeneous business environments.

A recent research stream focuses on monitoring and responding to business situations detected through event patterns. For example SARI [3] provides an event-based rule management framework, which allows modelling business situations and exceptions with sense and respond rules. To wholly realize the potential of this research stream, it is essential to allow business users to model and intuitively comprehend the appropriate responses to business situations by using concepts that are familiar to them, like milestones and goals. Goal-orientation is based on separating the declarative statements defining desired system behaviour from the various ways to achieve that behaviour, thus hiding from business users the details about low-level events.

Hence, there is a clear need for an event-based goal-oriented framework, which would provide recommendations for reacting to interesting or critical business situations, while it would increase the awareness of business users regarding the running business processes.

In this paper, we focus on the challenges of enabling awareness about the business situations happening in business environments and increasing organizations' responsiveness to them. We present BEAM, an event-driven framework for BusinEss Awareness Management, which aims to manage, i.e. monitor and control over time business situations and business systems that support the execution of business processes.

## 2. Related Work

To enable Business Awareness Management, we examined and reviewed technologies that can enable awareness about changing circumstances that may require reactions as well as mechanisms for monitoring business activities. In this context, we consider relevant for our work approaches from the research fields of Situation Awareness, Business Activity Monitoring (BAM), goal-orientation and Complex Event Processing (CEP).

*Situation awareness* was introduced by Mica Endsley whose definition of the term is a generally accepted one: "Situation Awareness is the perception of the elements in the environment within a volume of time and space, the comprehension of their meaning, and the projection of their status in the near future" [4]. Since this original work, a lot of situation-related research has been carried out and has become a critical issue in domains in which there is the need to automatically and continuously identify and act on complex, often incomplete and unpredictable, dynamic situations; as a result, effective methods of situation recognition, prediction, reasoning, and control are required — operations collectively identifiable as situation management (SM) [5].

*BAM* describes the processes and technologies that provide real-time situation awareness, along with access to, and analysis of, the critical business performance indicators, based on the event-driven sources of data [6]. BAM is used to improve the speed and effectiveness of business operations by keeping track of what is happening now, and raising awareness of issues as soon as they can be detected. BAM applications may emit alerts about a business opportunity or problem, drive a dashboard with metrics or status, make use of predictive and historical information, display an event log, and offer drill-down features [7].

In principle, *goal-orientation* is based on separating the declarative statements defining desired system behaviour from the various ways to achieve that behaviour [8]. Rimassa and Burmeister [9] propose GO-BPMN, a visual modelling language for the specification of business processes, which is an extension of the OMG standard BPMN. This notation helps to add goals, activity plans, and their relationships to process models. The Tibco's approach for goal-driven BPM [10], [11] follows a process of sense-and-response incremental improvements, making possible the creation of the most dynamic, agile, and responsive processes. For the "sense" part of the aforementioned process, the approach exploits CEP in order to identify important business-worthy events and respond to them; i.e. the response is event-driven.

*CEP* is a very active field of research and is being approached from many angles. A multitude of languages are proposed to formulate complex event patterns and different event processing paradigms are proposed to match these patterns over events [2], [12].

The framework proposed in the context of this paper enhances existing BAM applications by incorporating concepts and technologies from the research areas of CEP, situation awareness and goal-oriented business modelling. According to [13], most key performance indicators (KPIs) in business activity monitoring and performance management scenarios are complex events (although not all complex events are KPIs). Although most process monitors implement a basic form of CEP in the sense that they apply rules and perform computations on multiple event objects to calculate what is happening in a business process, they are not general-purpose CEP engines, and they don't "listen" to events from outside the managed business process. The proposed approach links

CEP with business process monitoring, allowing the provision of a broad, robust situation awareness capability that encompasses both internal process events and external business events. BAM on the basis of CEP shall improve the existing, often complained IT blindness, which is caused by thousands of low-level events per second without any semantics [14].

BAM usually sets up target values for each performance indicator. These target values are not the goals of the business processes, but the goals of the performance indicators. They usually lack meaning or purpose and are just values to be reached. They exist separately without relationships and hence it is hard to share a united vision of the monitored business processes. Therefore, there is a need to align the business processes with strategic goal architectures, which requires a change in emphasis from process to goal-oriented monitoring [15]. The proposed BEAM layer enables goal-oriented monitoring by using a process description oriented toward goals related to interesting situations, instead of using BPMN (or similar) process descriptions; for an example of the later approach please see [16], where the authors present a general framework for edBPM as well as a use case in the context of a large logistic company.

## 3. Approach and Architecture

Taking into account the need of business users to be continuously aware about what's happening in the business environment, as well as the need to adapt business processes according to the current business situations, BEAM aims to: a) notify business users about interesting or critical situations in the business environment and provide recommendations for reactions, b) recommend dynamically to business users new situations and events to be monitored in order to allow them to 'get the meaning' of the current business situations and acquire meaningful detailed information that enhance their current awareness regarding the running business processes and c) trigger adaptations of the running business processes by calling relevant services in an event-driven SOA setting where BEAM is part of.

BEAM is based on the assumption that there exist specific goals that a business process and system should fulfil and proposes the adoption of a goal-directed model able to track the fulfilment of goals at run time. We utilize goal-directed modelling in which we follow a hierarchical goal decomposition model we have developed as part of our previous work, called Situation-Action-Network (SAN) [17]. In SANs, goals are related to situations that trigger their activation and reactions that should be performed towards achieving goals if certain conditions are met; for details about the SAN models the reader is referred to [18].

In BEAM, the detection of critical and/or interesting situations is performed by employing CEP capabilities. BEAM exploits the complex event patterns identified by the CEP engine used (see [19]) in order to sense critical/interesting situations. Then, by taking into account the current business context, BEAM recommends appropriate responses with the aim to cope with problematic situations or exploit opportunities that may arise in the business environment.

The BEAM layer consists of several sub-components allowing the definition of goal-oriented situation-aware recommendations, the modelling of desired, meaningful reactions to interesting situations and the execution of business awareness monitoring (Figure 1). The SAN Editor is a graphical design tool developed in Adobe Flash/Flex. It is used for the development of SAN models represented in RDF (Resource Description Framework). SAN Engine undertakes the traversing of SAN trees stored in SAN Repository (implemented using Sesame 2.6.2) through the Traversal Service, as well as the subscription/unsubscription of complex event patterns in the Pub/Sub through the Subscriptions Management subcomponent. The Context Management subcomponent updates the current context and evaluates the necessary contextual conditions based on detected situations. Context changes are stored in the Context Repository. Action Service triggers actions in external systems and recommends actions to human actors by transmitting events through the Pub/Sub. Finally, the SAN engine is responsible for recommending actions which include notifications to users, subscriptions to other simple or complex events, and adaptations to running business processes.

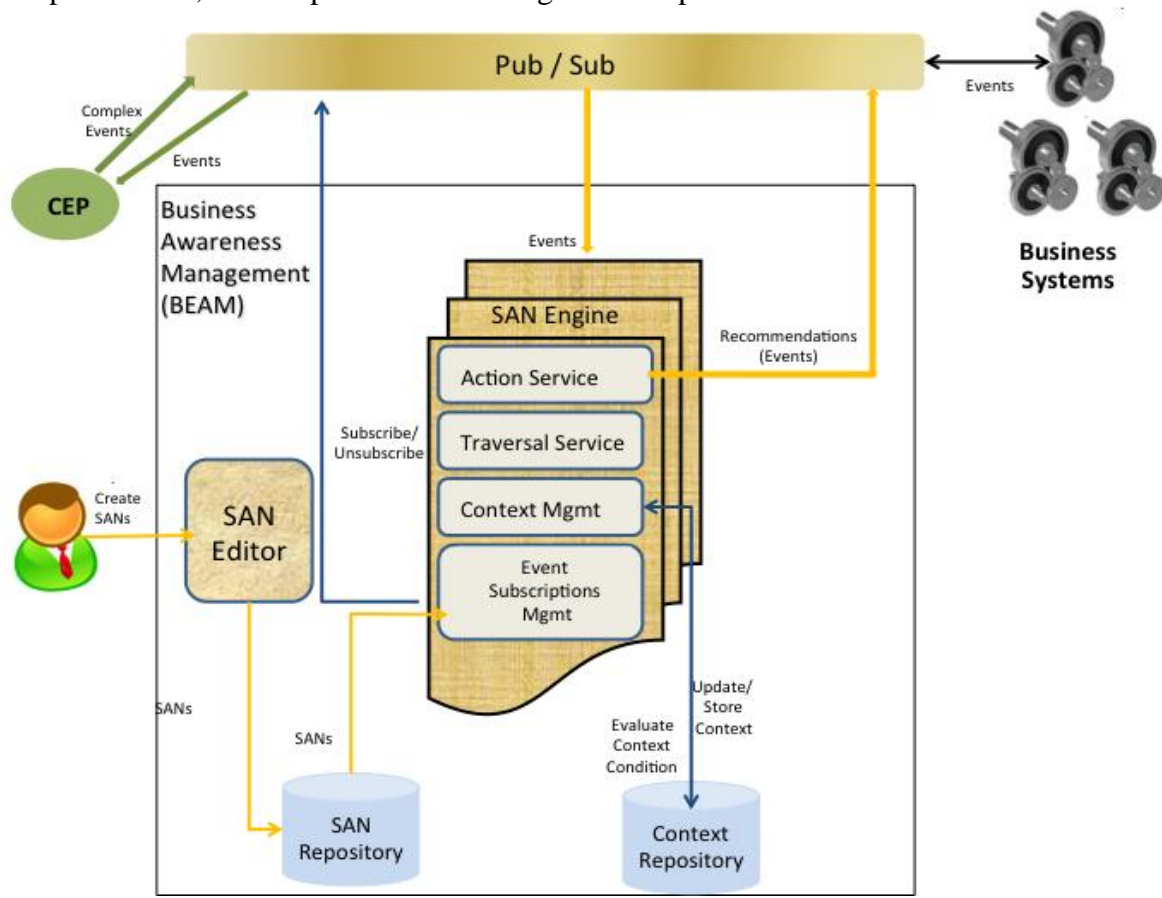

*Figure 1: BEAM Architecture*

## 4. A Business Awareness Monitoring Application

In the section we present a practical application of the proposed BEAM framework in the transportation and logistics domain and more specifically in a fast moving consumer goods (FMCG) SME. In order to understand how BEAM can increase the awareness of managers about the running business processes, as well as their flexibility we discuss in the following an indicative pilot scenario. The pilot scenario is then used for describing the practical role and use of the BEAM framework. The scenario includes the IT infrastructure in the headquarters of the company, comprising of several centralized corporate systems such as customer relationship management system (CRM), warehouse management system (WMS) and geographical information system (GIS), which are required for the management of orders and warehouse as well as the monitoring of trucks respectively.

### 4.1 Pilot Scenario

A typical process in the daily business of the FMCG SME is the order process. The as-is situation before adopting BEAM is as follows. A type of the FMCG company's employees, the preseller visits clients regularly (e.g. twice per week). Some of the clients order products and the order is entered by the preseller's mobile phone into the CRM system. The warehouse manager is notified about the order through the CRM system and proceeds with the appropriate actions for collecting and preparing the order (e.g. fill the palettes). One of these actions is to prepare the touring (routing) plan for the trucks. Presellers and trucks have predefined routing plans. The routing plans of the trucks are a subset of the preseller's routing plans, in the sense that the truck delivers the orders only to those customers that issued them. Usually trucks follow presellers with a time lag of one day. Finally one important legal restriction is that in case a truck is loaded with products the driver should possess all necessary legal documents (e.g. order form, invoice etc.) before starting transportation towards the customer.

Some customers may want to return parts of an order, even days after the order has been delivered. A customer that wants to return an order can either call the preseller and inform him about his intention, or can wait for his next normal visit to occur. In both cases the customer fills in the so-called returnee form, which is approved and signed by the preseller in two copies. The preseller just signs the copies and continues his planned route. Both copies remain to the customer. At the next itinerary, the truck driver sees the signed copy and loads the returnees (returned products), which go back to the warehouse together with the signed copy.

In the current state there are several problems related to the order return process (see Figure 2). The customer has to wait until the next planned itinerary of the truck driver in order to return the products. In the meanwhile these products decrease customer's warehouse capacity by occupying space or even worse they may expire. Moreover, while several drivers of the company having available space in their trucks may pass nearby the customer on their way to other itineraries, both the drivers and the warehouse manager lack the necessary knowledge to take the right decision, i.e. to stop to the customer with the returnees. These problems result in delays, additional costs for the FMCG SME in terms of personnel, transportation and wastage, unsatisfied customers and even lost customers in the very competitive FMCG industry, where speed is becoming the main competitive weapon [20].

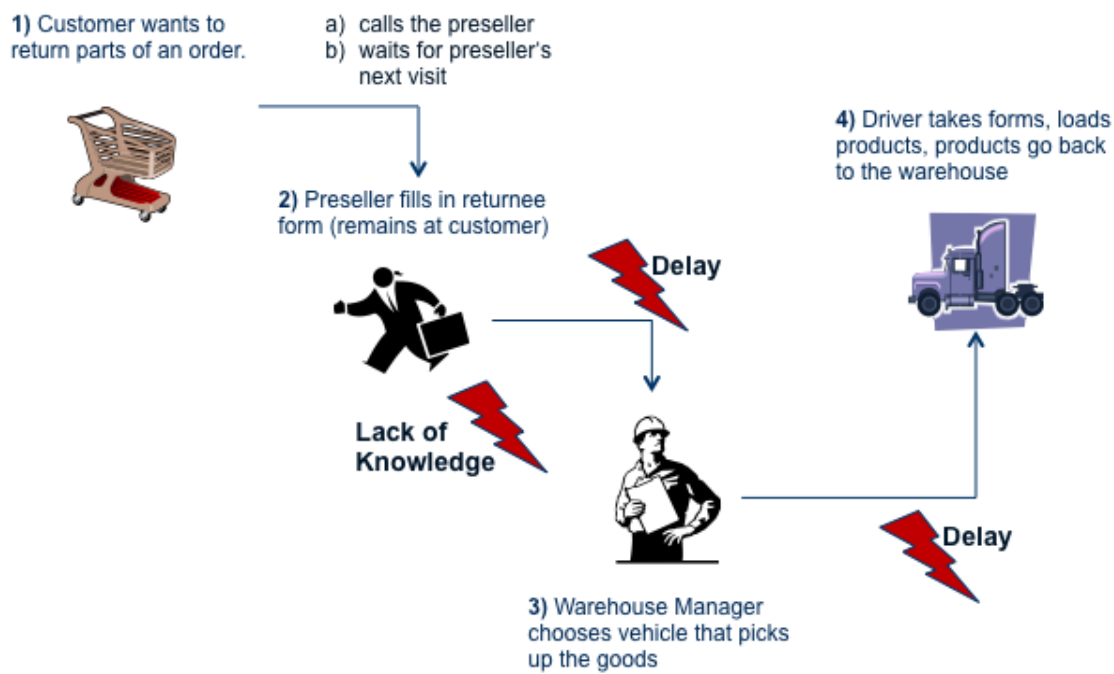

*Figure 2: As-is state of order return scenario*

*4.2  System Walkthrough*

The BEAM architecture in the context of the order return scenario is depicted in Figure 3. Information about new orders, returnee requests as well as other data related to company's interactions with customers, clients, and sales prospects are stored into FMCG SME's CRM. A mobile CRM client allows presellers to update remotely the CRM system though their mobile device with information about returnees, orders and customers. WMS is a key part of the supply chain and primarily aims to control the movement and storage of materials within FMCG company's warehouse and process the associated transactions, including shipping, receiving, putaway and picking. The tracker devices along with the fuel, battery and other sensors, which have been mounted into the FMCG company's vehicles, provide continuous information about vehicles' position, speed, fuel level, state etc. Events from the various information sources (CRM, WMS and trucks) are communicated to the Pub-Sub broker through the respective CRM, WMS and GIS adapters, which are responsible for transforming information entered into the corresponding databases to an appropriate event format. Finally, the events created are pushed to the intermediary Pub/Sub broker, while subscribers register subscriptions with that broker under predefined subscription terms, letting the broker perform the filtering.

Consider the situation where a returnee request has been delivered to the company, while in the same time window (e.g. within 2-3 hours) one of the company's trucks is entering a geographic area, which is rather close to the customer who wants to return the products. This may be an opportunity for the company, as it may be beneficial to add dynamically an additional stop to the truck's predefined routing plan. In this scenario BEAM investigates the feasibility and benefits of the various alternative actions and recommends the best one, while it also informs the managers about the detected situation(s).

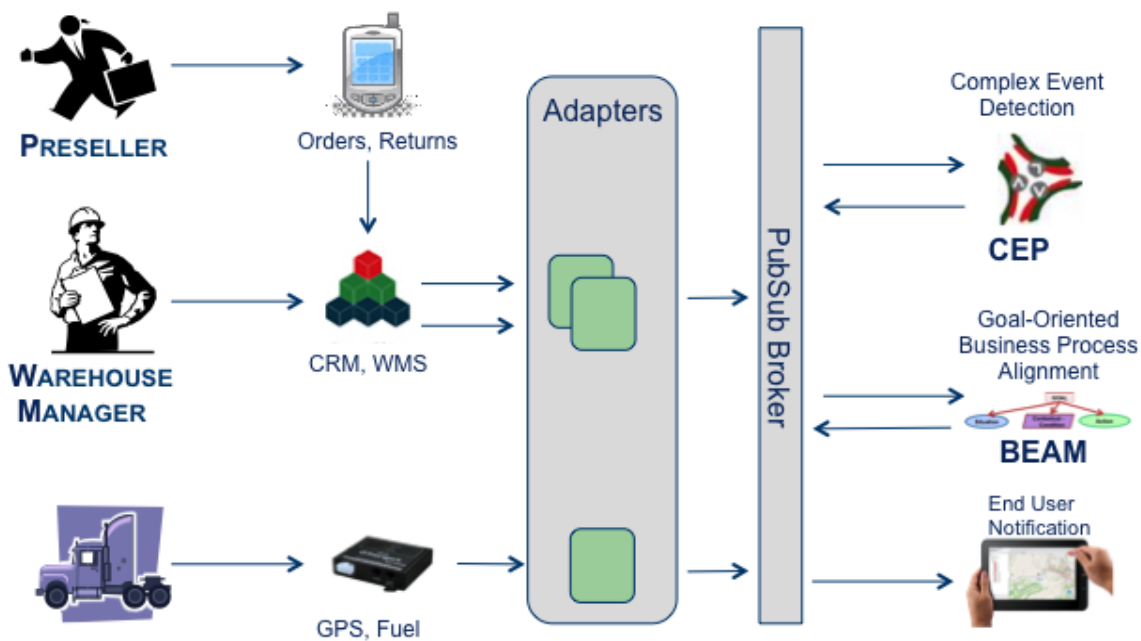

*Figure 3: Overall Architecture*

First, BEAM by following the underlying SAN model subscribes to the complex event pattern '*ExtraStopOpportunity*', which corresponds to the situation described above and is detected by using the Etalis CEP engine [19] by combining simple events from multiple sources (GPS, CRM, WRM). It should be noted that the modeller, who is also responsible for the definition of the SAN model, defines complex event patterns through an appropriate modelling tool. Second, it checks the current context. In some specific business contexts of the FMCG SME it makes no sense to even investigate the possibility to recommend the addition of the extra stop to the planned route. Examples include the case that the truck has not enough fuel, the case that the returnee request occurs close to the end of the company's business day, the case that the truck is failing to execute the deliveries according to the specified time schedule and so on. So, in all of the above cases BEAM makes no recommendation.

Otherwise, as a third step, BEAM proceeds with the recommendation or triggering of action(s) based on the detected situation and the current context. Actions could include a) manager notifications, b) triggering of other components of the event-driven SOA (e.g. GIS) with the purpose to perform specific functionalities (e.g. reroute a truck, estimate the time needed to approach a customer etc.) and c) subscription to other complex events with the purpose to get additional information that will be used in order to find out the appropriate reaction. In this scenario some possible BEAM actions are the following:

- To trigger GIS (type b action – see above) by publishing an appropriate event to the Pub-Sub (where GIS has been subscribed) so that GIS can indicate on the display of the driver the addition of an extra stop.
- After that, to inform the warehouse manager, e.g. through an RSS feed, that a new stop has been added to the truck's route (type a action).
- Finally, an example of type c action in another, or even in the same, scenario is the following: in case a truck moves out of a predefined geographical zone, BEAM will recommend the subscription to fuel consumption events, in order to increase manager's awareness about the current situation and ensure that the truck will not run out of fuel.

## 5. Conclusions and Future Work

Modern organizations need real-time awareness about the current business conditions and the various events that occur from multiple and heterogeneous environments. Moreover, the need for flexible processes is big in today's environment, especially for SMEs as, unlike larger companies, they can't lean on globally recognized brands: a lost customer, or a missed opportunity to recruit a new customer, may never be recouped. With limited resources, SMEs must struggle to adapt to changing market conditions, whether these are reduced time to market, increased expected service levels, or price-cutting.

In this paper we present BEAM, a software framework dealing with such a need. A preliminary evaluation of BEAM has shown that user feedback has been positive and confirms that the system is in-line with our primary goal to increase the flexibility of loosely coupled business systems. More specifically, the managers of the logistics company that used the system reported that the event-driven flexibility introduced by BEAM leads to constant improvement of the process of value creation and delivery through minimizing costs and work errors and raising the satisfaction of clients and other stakeholders. A more complete evaluation of the proposed framework will be done within an FP7 SME project until the end of the year.

## Acknowledgement

This work is partly funded by the European Commission projects FP7 SME ReFLEX (262305) and FP7 ICT PLAY (258659).